\documentclass[11pt,a4paper]{article}
\usepackage[T1]{fontenc}
\usepackage{lmodern}
\usepackage[margin=20mm]{geometry}
\usepackage{microtype}
\usepackage{amsmath,amssymb,bm}
\usepackage{graphicx}
\usepackage[export]{adjustbox}
\usepackage{etoolbox}
\AtBeginEnvironment{table}{\small\centering}
\usepackage{booktabs}
\usepackage{siunitx}
\usepackage{xcolor}
\usepackage[font=small,labelfont=bf,labelsep=period]{caption}
\usepackage[numbers,sort&compress]{natbib}
\usepackage{xurl}
\usepackage{hyperref}
\definecolor{citelink}{RGB}{0,90,160}
\hypersetup{colorlinks=true,citecolor=citelink,linkcolor=black,urlcolor=citelink,
  pdftitle={SPECTRA-RCWA: A Spectral Propagation Engine for Coupled-Wave Theory with Recursive Admittance},pdfauthor={Hyoseok Park}}
\graphicspath{{figs/}}

\newcommand{\mat}[1]{\mathbf{#1}}

\title{\Large\bfseries SPECTRA-RCWA: A Spectral Propagation Engine for Coupled-Wave Theory with Recursive Admittance}
\author{Hyoseok Park\\[6pt]
\small Department of Physics and Institute of Quantum Systems,\\
\small Chungnam National University, Daejeon 34134, Republic of Korea\\
\small \href{mailto:phs137@o.cnu.ac.kr}{phs137@o.cnu.ac.kr}}
\date{}

\begin{document}
\maketitle
\begin{abstract}
Rigorous coupled-wave analysis (RCWA) is the standard full-wave solver for periodic metasurfaces, but every solve diagonalises a dense nonsymmetric matrix for each layer. Inverse design repeats this step for every wavelength, angle and polarisation of an objective. On a graphics processor it is the slow step, and its derivative fails where eigenvalues coincide or a diffraction order reaches cutoff. Here we introduce SPECTRA-RCWA, an RCWA forward and adjoint solver that never diagonalises the layer matrix. Each layer step uses three entire functions of the half-size layer matrix, evaluated by matrix products alone, and an admittance recursion, or scattering-matrix doubling for thick layers, keeps the evanescent orders stable. One adjoint field gives the gradient for every design value without a computational graph, and all solves of an objective run as one batch. For 144 solves with 578 $\times$ 578 layer matrices, the value and the gradient take 0.25 s on one graphics card, 218 times less than with a differentiable eigendecomposition code. The truncated-series solver TorchRDIT loses accuracy beyond a few tenths of a micrometre unless the user slices the layer; SPECTRA-RCWA keeps single-precision accuracy in layers several wavelengths thick with no setting changed and is 47 to 65 times faster at matched accuracy. Its single-precision gradient stays within $4\times10^{-5}$ of a double-precision reference at a cutoff, where eigendecomposition gradients are wrong by order one. A freeform metagrating designed in 3.4 min on a desktop processor deflects both polarisations by 65$^\circ$ with a band-averaged efficiency of 75\%.
\end{abstract}

\section{Introduction}

Freeform metasurfaces are designed by gradient-based optimisation over thousands of design values, with the gradient taken from an adjoint simulation or from automatic differentiation.\cite{molesky2018inverse,jensen2011topology,lalaukeraly2013adjoint,hughes2018adjoint,christiansen2021inverse,piggott2015inverse} For periodic structures such as metagratings, the unit cells of large-area metasurfaces and the pixels of colour-routing image sensors, the field solver of choice is rigorous coupled-wave analysis (RCWA), also called the Fourier modal method.\cite{sell2017large,phan2019high,jiang2019global,lin2019topology,pestourie2018inverse,mansouree2021large,miyata2021full,zou2022pixel,catrysse2022subwavelength} The objective of such a device is rarely one solve. A broadband device under a lens needs every wavelength, every incidence angle of the illumination cone and both polarisations, tens or hundreds of solves per evaluation, and an optimisation repeats the evaluation hundreds of times.

RCWA expands the fields of each layer in Fourier orders and propagates them through the layer with the eigenvectors and eigenvalues of a dense, nonsymmetric layer matrix.\cite{moharam1981rigorous,moharam1995formulation,li1997new,lalanne1997improved} Decades of work have made the method converge quickly, through the Fourier factorisation rules,\cite{li1996fourier,lalanne1996highly,granet1996efficient,popov2000grating,popov2001maxwell} and made it stable, through transmittance-matrix and scattering-matrix recursions.\cite{moharam1995stable,li1996formulation,whittaker1999scattering,rumpf2011improved,liu2012s4} The eigendecomposition itself has stayed, and it limits gradient-based design in two ways. On a graphics processor it is the slow step. The reduction of a nonsymmetric matrix to Hessenberg form is a largely sequential computation that is shared with the host processor,\cite{tomov2010accelerating} while matrix products and factorisations run efficiently as batches,\cite{abdelfattah2016performance,dongarra2017design} and differentiable RCWA codes\cite{kim2023torcwa,jin2020inverse,colburn2021inverse,schubert2023fourier} spend a large part of every solve in this step. Its derivative is also singular at two kinds of points. The derivative of the eigenvectors contains the factors $1/(\lambda_i-\lambda_j)$ of every pair of eigenvalues\cite{giles2008collected,vanderaa2007computation} and is regularised in practice by a broadening parameter.\cite{liao2019differentiable,colburn2021inverse,kim2023torcwa} At an eigenvalue of zero, where a diffraction order of the layer passes from propagating to evanescent, the square root that defines its propagation constant has no derivative.

Several formulations avoid the eigenproblem. The differential method integrates the coupled-wave equations numerically through the layer.\cite{popov2001maxwell,neviere2018light} Rigorous diffraction interface theory replaces the modal solution of a thin layer by a truncated power series,\cite{roberts2017rigorous} and TorchRDIT brought this series to graphics processors and to inverse design with automatic differentiation.\cite{huang2024eigendecomposition} Other authors approximate the layer exponential by a polynomial\cite{xu2024highly} or compute it from a matrix square root.\cite{vanderceelen2026matrix} The derivative of a modal scattering matrix has been written without eigenvector derivatives,\cite{zhu2020differentiable} and perturbative corrections reduce the number of eigendecompositions for structures that vary along the propagation direction.\cite{zhu2022varrcwa} A series of fixed order is accurate only up to a thickness that the user has to find, and a gradient by automatic differentiation keeps every term of the series in memory. In numerical analysis, the functions that propagate a second-order system, the matrix cosine and its relatives, are computed to rounding accuracy by scaling and double-angle recurrences.\cite{serbin1980algorithm,almohy2015algorithms,nadukandi2018computing,higham2008functions}

Here we introduce SPECTRA-RCWA (spectral propagation engine for coupled-wave theory with recursive admittance), an RCWA solver that returns the value and the gradient of an objective for all its wavelengths, angles and polarisations in one batch of matrix products and inverses. A layer step is written with three entire functions of the half-size layer matrix, which we evaluate by scaling, a Paterson--Stockmeyer polynomial and double-angle steps. The step is accurate to rounding at any thickness, and nothing in it is singular at an order at cutoff. Stability comes from an admittance recursion with one pivoted inverse per step and, for thick layers, from scattering-matrix doubling in the basis of a lossy reference medium. The gradient comes from one adjoint field that is derived by hand, with no computational graph and no division by eigenvalue differences or square roots.

We measure accuracy and speed against S4, torcwa, a batched eigendecomposition code and TorchRDIT, and we use the solver to design a freeform metagrating on a desktop processor. An evaluation that takes a differentiable eigendecomposition code about a minute takes a quarter of a second, which turns an optimisation of a thousand iterations from an overnight computation into one of four minutes. Against TorchRDIT, the closest published solver, the difference lies in error control and in the cost of the gradient. Its series loses accuracy beyond a few tenths of a micrometre unless the user slices the layer, and the step rule of SPECTRA-RCWA holds single-precision accuracy up to \SI{3}{\micro\metre} with no setting changed. With the gradient taken from one adjoint field rather than from a graph of the series, the value and the gradient are 47 to 65 times faster on one graphics card at matched accuracy.

\section{Formulation}

\subsection{Layer Step from Entire Functions}

\begin{figure}
\centering
\includegraphics[max width=\textwidth]{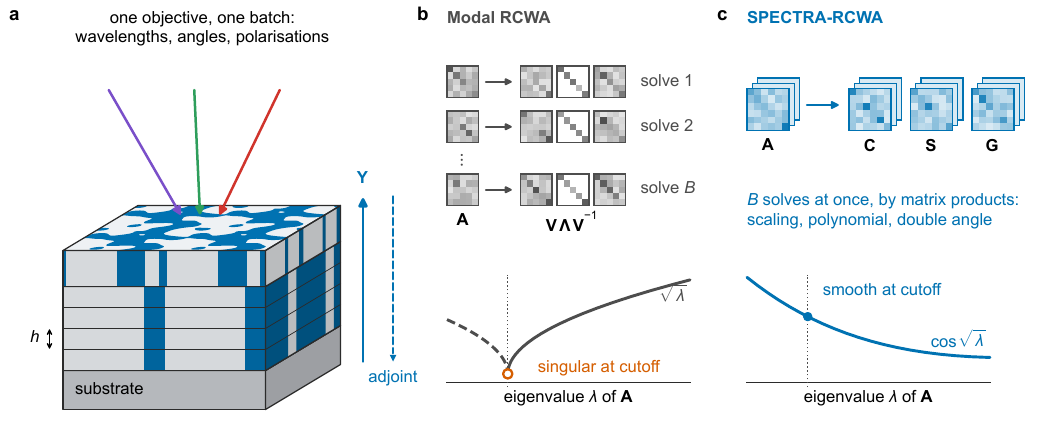}
\caption{\textbf{SPECTRA-RCWA replaces the eigendecomposition of each layer by matrix functions of the whole batch.} (a) A stack of periodic layers on a substrate, illuminated by all wavelengths, angles and polarisations of an objective at once. Each layer is cut into steps of thickness $h$. The admittance $\mat Y$ is carried from the substrate to the top, and one adjoint field runs back through the same steps and gives the gradient. The pattern is synthetic. (b) Modal RCWA diagonalises the layer matrix $\mat A$ of each of the $B$ solves in turn, $\mat A=\mat V\mat\Lambda\mat V^{-1}$. The propagation constant of a mode is the square root of its eigenvalue, which has no derivative at zero, where an order is at cutoff (dashed line: imaginary part). (c) SPECTRA-RCWA forms the entire functions $\mat C$, $\mat S$ and $\mat G$ of $\mat A$ for the whole batch by matrix products. The cosine of the square root, the function behind $\mat C$, is smooth through the cutoff.}
\label{fig:concept}
\end{figure}

A layer step can be written with functions of the layer matrix that have no singular point. We consider a stack of layers that are periodic in $x$ and $y$ and uniform along $z$ within each layer (Figure~\ref{fig:concept}a). With $N$ Fourier orders retained, the tangential electric and magnetic fields of a layer are vectors $\mat e(z)$ and $\mat h(z)$ of length $2N$ that obey
\begin{equation}
\frac{d\mat e}{dz}=i\omega\,\mat P\,\mat h,\qquad \frac{d\mat h}{dz}=i\omega\,\mat Q\,\mat e ,
\label{eq:ode}
\end{equation}
where $\omega=2\pi/\lambda$ is the vacuum wavenumber and the $2N\times2N$ matrices $\mat P$ and $\mat Q$ hold the transverse wavevectors, in units of $\omega$, and the Fourier matrix of the permittivity (Supplementary Information, Section S1). Eliminating $\mat h$ gives $d^2\mat e/dz^2=-\omega^2\mat A\,\mat e$ with the layer matrix $\mat A=\mat P\mat Q$. Modal RCWA diagonalises $\mat A$: the eigenvalues are the squared propagation constants of the layer modes and the eigenvectors are their field profiles (Figure~\ref{fig:concept}b).

A second-order system with constant coefficients is propagated by a cosine and a sine of the square root of its matrix. For a step of thickness $h$ we define
\begin{equation}
\mat C=\cos\!\big(\omega h\mat A^{1/2}\big),\qquad
\mat S=\sin\!\big(\omega h\mat A^{1/2}\big)\,\mat A^{-1/2},\qquad
\mat G=(\mat C-\mat I)\,\mat A^{-1},
\label{eq:csg}
\end{equation}
and the fields at the top of the step, $z=0$, follow from those at its bottom as
\begin{equation}
\mat e(0)=\mat C\,\mat e(h)-i\,\mat S\mat P\,\mat h(h),\qquad
\mat h(0)=-i\,\mat Q\mat S\,\mat e(h)+(\mat I+\mat Q\mat G\mat P)\,\mat h(h).
\label{eq:step}
\end{equation}
The three functions of eq~\ref{eq:csg} are power series in $\mat A$ itself; the series of $\mat C$, for instance, has the terms $(-1)^k(\omega h)^{2k}\mat A^k/(2k)!$. They contain no square root and no inverse, and they are entire: they have a derivative at every $\mat A$, including a singular one. An order at cutoff, which corresponds to an eigenvalue of zero, is an ordinary point of eq~\ref{eq:step}, and neither $\mat P$ nor $\mat A$ is inverted anywhere.

The series cannot simply be truncated. It is accurate for thin layers and loses digits quickly beyond them, because its terms grow before they decay. We use instead the method of numerical analysis for these functions.\cite{serbin1980algorithm,almohy2015algorithms,nadukandi2018computing} The argument $(\omega h)^2\mat A$ is divided by $4^s$, with $s$ chosen from its norm. A polynomial of degree six in single precision and eight in double precision gives $\mat C-\mat I$ and $\mat S$ of the scaled argument by the Paterson--Stockmeyer scheme,\cite{paterson1973number} with four matrix products in single precision, and $s$ double-angle steps restore the full argument. The steps act on $\mat C-\mat I$, which keeps the digits of a small argument, and $\mat G$ follows from the half-angle identity $(\cos x-1)/x^2=-\tfrac12[\sin(x/2)/(x/2)]^2$ with one more product. The whole evaluation consists of matrix products, the operation that a graphics processor runs fastest and that batches over solves without any change. The structure of $\mat P$ and $\mat Q$ gives $\mat A$ from two products of a quarter of the full size (Supplementary Information, Section S1).

\subsection{Admittance Recursion and Doubling}

An admittance recursion keeps the evanescent orders stable without modes. Such an order grows as $\cosh(\omega h k)$ in $\mat C$ and $\mat S$, so eq~\ref{eq:step} cannot be applied across a thick layer in floating-point arithmetic. Scattering-matrix algorithms solve this problem in the modal basis by separating growing and decaying modes.\cite{li1996formulation,whittaker1999scattering,liu2012s4} We cut each layer into steps with $\omega h k_{\max}\le4.5$, where $k_{\max}$ is the largest retained transverse wavevector, so that no entry of the step exceeds about 45, and we carry the admittance matrix $\mat Y$, defined by $\mat h=\mat Y\mat e$, upward from the output medium. With $\mat h(h)=\mat Y\mat e(h)$ below a step, eq~\ref{eq:step} gives
\begin{equation}
\mat D=(\mat C-i\,\mat S\mat P\mat Y)^{-1},\qquad
\mat e(h)=\mat D\,\mat e(0),\qquad
\mat Y_{\rm top}=(\mat Y+\mat Q\mat G\mat P\mat Y-i\,\mat Q\mat S)\,\mat D .
\label{eq:admittance}
\end{equation}
One pivoted inverse per step is the only operation besides products. The admittance stays bounded because it describes a passive half-space, and the recursion runs unchanged through any number of layers. At the top of the stack, the admittance and that of the incidence medium give the reflected field, and the saved factors $\mat D$ carry the field back down to the substrate for two right-hand sides, one per polarisation. Recursions of this kind for an admittance or impedance matrix are well established for acoustic waveguides,\cite{pagneux1996study} where they also serve to control evanescent modes.

The number of steps grows in proportion to the thickness. For a layer that needs seven steps or more we build its scattering matrix by doubling, as in the adding and doubling methods of radiative transfer and the precise integration method.\cite{hansen1974light,zhong2004precise} One thin step of eq~\ref{eq:step} is written in the wave basis of a reference medium as a transmission matrix $\mat T$ and a reflection matrix $\mat R$, which are identical from both sides because the layer is uniform. Joining the step to itself by the Redheffer star product\cite{redheffer1962relation} doubles the thickness,
\begin{equation}
\mat T_2=\mat T(\mat I-\mat R\mat R)^{-1}\mat T,\qquad
\mat R_2=\mat R+\mat T(\mat I-\mat R\mat R)^{-1}\mat R\,\mat T ,
\label{eq:doubling}
\end{equation}
so that $r$ doublings give $2^r$ steps with $r+1$ inverses. The reference medium is a fictitious homogeneous one with the lossy permittivity $4+4i$. In a vacuum basis, $\mat I-\mat R\mat R$ becomes singular at a guided mode of the layer, and the basis itself degenerates when an order is at grazing incidence; with a lossy reference neither occurs for real wavevectors. The scattering matrix of a layer depends on that layer alone, so it can be stored and reused, for example for a fixed layer stack below a design.

\subsection{Gradient from One Adjoint Field}

One adjoint field gives the derivative with respect to every design value of every layer. A design changes the permittivity of a layer and through it $\mat P$ and $\mat Q$. For an objective $J$ that depends on the fields leaving the stack, the first-order change is
\begin{equation}
\delta J=\operatorname{Re}\int i\omega\left(\mat a^{H}\,\delta\mat P\,\mat h+\mat b^{H}\,\delta\mat Q\,\mat e\right)dz ,
\qquad
\frac{d\mat a}{dz}=i\omega\,\mat Q^{H}\mat b,\quad \frac{d\mat b}{dz}=i\omega\,\mat P^{H}\mat a ,
\label{eq:adjoint}
\end{equation}
where the adjoint field $(\mat a,\mat b)$ obeys the transposed system, is driven at the bottom of the stack by the derivative of $J$ with respect to the transmitted field and satisfies the radiation condition of the incidence medium at the top (Supplementary Information, Section S2). Equation~\ref{eq:adjoint} is the adjoint method of electromagnetic design\cite{lalaukeraly2013adjoint,hughes2018adjoint} written for the Fourier amplitudes of a layer.

The adjoint field needs no second factorisation. The factors $\mat D$ of eq~\ref{eq:admittance}, or the doubling tree of eq~\ref{eq:doubling}, give $(\mat e,\mat h)$ and $(\mat a,\mat b)$ on the planes between the steps in products with two columns, each in the direction in which it is stable. Between the planes the integrand is smooth, and a short Taylor series in $\mat P$ and $\mat Q$ evaluates it on Gauss--Legendre nodes. The integral becomes a sum of low-rank terms that is mapped to the design values through the Fourier transform of the pattern. No intermediate matrix of the forward pass is kept for a computational graph, and eq~\ref{eq:adjoint} contains neither an eigenvalue difference nor the square root of an eigenvalue. The gradient error near degenerate or vanishing eigenvalues is set by the precision of the arithmetic and not by the distance to the degeneracy. Any formulation that differentiates a power series shares this property, TorchRDIT included, and it separates both from formulations that differentiate an eigendecomposition or a matrix square root.

\subsection{One Batch per Objective}

Every operation above is a matrix product, a pivoted inverse or an elementwise combination, and each applies to a stack of matrices as well as to one. All solves of an objective, for every wavelength, incidence direction and both polarisations, are stacked along one batch dimension and advance through the layers together. The two polarisations share all matrices and differ only in the two right-hand sides. Memory is the single resource that limits the batch, and a larger objective is split into chunks that fit.

\section{Results}

SPECTRA-RCWA reproduces the modal solution to rounding error and keeps its gradient accurate where modal gradients fail. In the following, we test the layer step and the complete solver, the gradient next to a cutoff, the comparison with TorchRDIT and the speed of a full objective on a graphics processor, and we then design a metagrating.

\subsection{Accuracy of the Layer Step and of the Solve}

\begin{figure}
\centering
\includegraphics[max width=\textwidth]{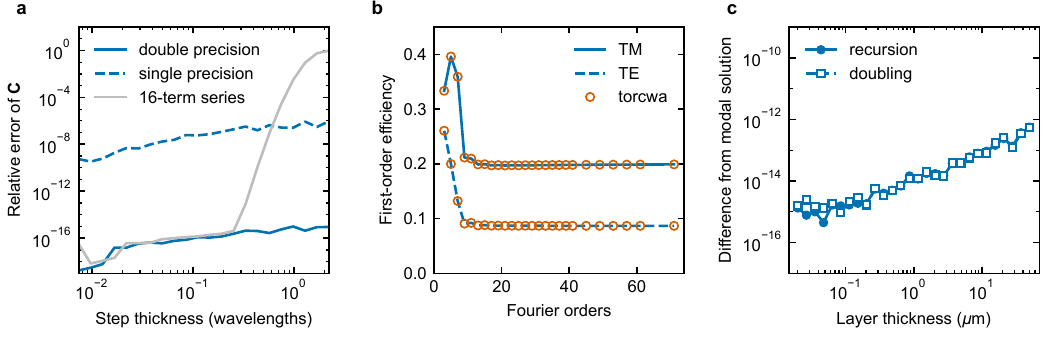}
\caption{\textbf{The layer step is accurate to rounding, and the complete solver reproduces the modal solution.} (a) Relative error of $\mat C$ against a 256-bit reference as a function of the step thickness, for the scaled evaluation in double and single precision and for a 16-term power series without scaling. The layer matrix is that of a grating of index 2.0 with 25 orders at a wavelength of 540 nm. (b) First-order transmission of a lamellar grating (index 2.0, period \SI{1}{\micro\metre}, depth \SI{0.5}{\micro\metre}, 633 nm, incidence at 10$^\circ$) against the number of Fourier orders, from SPECTRA-RCWA (lines) and torcwa (circles) in double precision. (c) Largest difference between the diffraction efficiencies of SPECTRA-RCWA and of a modal solution for a freeform layer of index 2.0 with 49 orders, against the layer thickness.}
\label{fig:accuracy}
\end{figure}

The scaled layer step is accurate to rounding at every thickness tested. For the layer matrix of a grating with 25 orders, we compare $\mat C$ with a reference computed in 256-bit arithmetic (Figure~\ref{fig:accuracy}a). With scaling and double-angle steps the relative error stays below $10^{-15}$ in double precision and below $10^{-6}$ in single precision for steps up to 2.2 wavelengths, the largest tested. A series of 16 terms without scaling matches the double-precision result up to a quarter of a wavelength, and its error exceeds $10^{-6}$ beyond 0.6 wavelengths. The other two functions behave in the same way (Figure S1).

The complete solver reproduces the modal solution. For a lamellar grating at oblique incidence the diffraction efficiencies of both polarisations agree with torcwa\cite{kim2023torcwa} in double precision to $1.2\times10^{-13}$ at every truncation from 3 to 71 orders, and the two formulations share one convergence curve (Figure~\ref{fig:accuracy}b). For a freeform layer with 49 orders, the difference from a modal solution stays below $6\times10^{-13}$ as the thickness grows from \SI{0.02}{\micro\metre} to \SI{50}{\micro\metre}, that is, to 79 wavelengths and 333 steps, for the recursion and for doubling alike (Figure~\ref{fig:accuracy}c). An absorbing layer gives differences below $10^{-14}$ (Figure S2).

\subsection{Gradient at a Cutoff}

\begin{figure}
\centering
\includegraphics[max width=\textwidth]{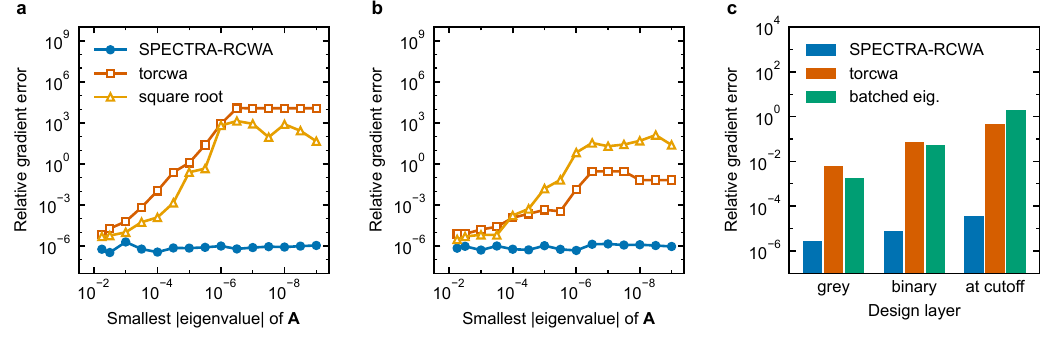}
\caption{\textbf{The adjoint gradient stays accurate next to a cutoff.} (a) Relative error of the single-precision gradient against the smallest eigenvalue of the layer matrix, at the cutoff where $\mat P$ becomes singular, for the adjoint field of SPECTRA-RCWA, for torcwa and for a square-root formulation. The layer is a freeform pattern of index 2.0, \SI{0.6}{\micro\metre} thick, with 49 orders and 256 design values, on a desktop processor. (b) The same at the cutoff where $\mat Q$ becomes singular. (c) Gradient error of the 144-solve benchmark on a graphics processor for three design layers, for SPECTRA-RCWA, torcwa and a batched eigendecomposition code. All errors are relative $L_2$ differences from a double-precision reference.}
\label{fig:gradient}
\end{figure}

The adjoint gradient keeps its accuracy at a cutoff, where gradients through an eigendecomposition fail. We tune the wavelength of a freeform layer so that the smallest eigenvalue of $\mat A$ approaches zero, and we compare single-precision gradients of the zeroth-order transmission with a double-precision reference. Two cutoffs exist for a two-dimensional pattern, one at which $\mat P$ becomes singular and one at which $\mat Q$ does. The adjoint field holds a relative error between $3\times10^{-7}$ and $2\times10^{-6}$ over the whole scan, from an eigenvalue of $6\times10^{-3}$ down to $10^{-9}$ (Figure~\ref{fig:gradient}a,b). The gradient of torcwa, which differentiates the eigendecomposition, starts near $10^{-5}$, exceeds 1\% at an eigenvalue of $10^{-4}$ on the first cutoff and saturates four orders of magnitude above the gradient itself. A square-root formulation implemented in our code, with a Denman--Beavers iteration\cite{denman1976matrix,higham1997stable} for $\mat A^{1/2}$ and a scaled Taylor exponential,\cite{higham2005scaling} follows a similar trend. In double precision the adjoint field agrees with reverse-mode differentiation of the same forward pass to $6\times10^{-15}$ and with central differences to $10^{-7}$ (Figure S3).

Near-cutoff layers are common in practice. In the benchmark of the next sections, a 17 $\times$ 17-order problem with 144 solves per evaluation, about one layer solve in 288 of a multilayer design has an inverse layer matrix with a norm above $10^{6}$, and such a solve can dominate the gradient of the whole evaluation. On the graphics processor, the single-precision gradient of SPECTRA-RCWA differs from the double-precision reference by $2.7\times10^{-6}$ for a grey design layer, $8.2\times10^{-6}$ for a binary one and $3.8\times10^{-5}$ for a layer with an order at cutoff. For these three layers torcwa gives $5.9\times10^{-3}$, $7.2\times10^{-2}$ and 0.49, and a batched eigendecomposition code gives $1.8\times10^{-3}$, $5.4\times10^{-2}$ and 2.0 (Figure~\ref{fig:gradient}c). For the grey layer the outputs of all three codes agree with the reference to better than $10^{-5}$. The loss is confined to the gradient, where an optimiser cannot detect it.

\subsection{Comparison with a Truncated-Series Solver}

\begin{figure}
\centering
\includegraphics[max width=\textwidth]{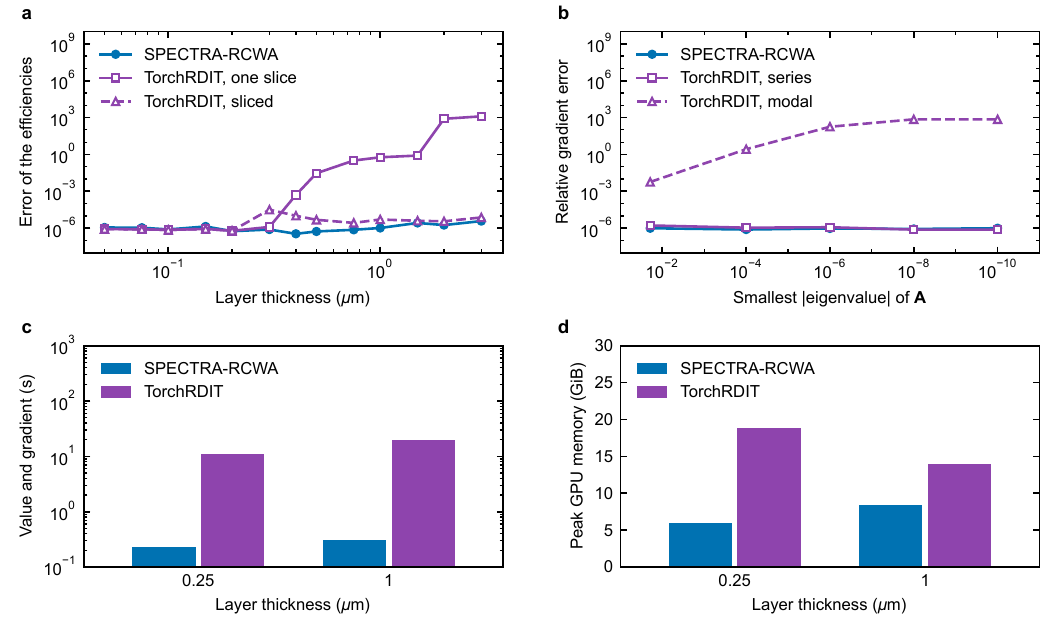}
\caption{\textbf{SPECTRA-RCWA controls its error without settings and returns the gradient faster and with less memory than TorchRDIT.} Comparison with TorchRDIT 0.2.0 in single precision. (a,b) Desktop processor, freeform layer of index 2.0 with 49 orders. (a) Largest error of the diffraction efficiencies against the layer thickness, for SPECTRA-RCWA, for TorchRDIT with one slice and series order 15, and for TorchRDIT with slices of \SI{0.25}{\micro\metre} and order 20. (b) Relative gradient error against the smallest eigenvalue of the layer matrix for SPECTRA-RCWA, for the series mode of TorchRDIT (order 20) and for its eigendecomposition mode. (c) Time for the value and the gradient of 144 solves with $578\times578$ layer matrices on one RTX 5090 card, with TorchRDIT at the settings that match the accuracy of SPECTRA-RCWA: series order 15 and one slice at \SI{0.25}{\micro\metre}, order 20 and slices of \SI{0.25}{\micro\metre} at \SI{1}{\micro\metre}. (d) Peak memory of the same runs. TorchRDIT took four (\SI{0.25}{\micro\metre}) and two (\SI{1}{\micro\metre}) of the 16 incidence directions per call to stay within 80\% of the card, SPECTRA-RCWA all 16.}
\label{fig:compare}
\end{figure}

Compared with TorchRDIT, SPECTRA-RCWA controls its error without user settings and computes the gradient in a fraction of the time and memory. TorchRDIT\cite{huang2024eigendecomposition} is the published solver closest to this work. It also avoids the eigendecomposition, by a power series of fixed order for each layer or for each slice of a layer that the user defines. We first compared the two codes on a desktop processor for a freeform layer with 49 orders, with the converged modal solution as the reference and each code in its own conventions. In single precision SPECTRA-RCWA holds an error between $3.5\times10^{-7}$ and $3.8\times10^{-6}$ for thicknesses from \SI{0.05}{\micro\metre} to \SI{3}{\micro\metre} with no setting changed (Figure~\ref{fig:compare}a). TorchRDIT with one slice and the series order of its paper reaches this accuracy up to \SI{0.3}{\micro\metre}; beyond that its error grows substantially, to order one. The energy balance does not reveal this: at \SI{0.75}{\micro\metre}, where one efficiency is wrong by 0.32, $T+R$ departs from one by $5\times10^{-5}$. With slices of \SI{0.25}{\micro\metre} set by hand and the largest order of its paper, TorchRDIT stays at about $3\times10^{-5}$ or below at every thickness. The two solvers differ in error control: the step rule and the scaling choose the work from the layer, where the truncated series needs the user to choose it.

Next to a cutoff both series-based gradients stay flat, between $7\times10^{-7}$ and $1.7\times10^{-6}$ in single precision down to an eigenvalue of $10^{-10}$. The eigendecomposition mode of TorchRDIT has a gradient error of $6\times10^{-3}$ away from the cutoff and of about 200 at an eigenvalue of $10^{-6}$ (Figure~\ref{fig:compare}b).

The two codes part most clearly in the cost of the gradient. We ran both on one RTX 5090 card for 144 solves with $578\times578$ layer matrices, the size of the benchmark of the next section, on a grey pattern of index 2.0. For a layer of \SI{0.25}{\micro\metre}, SPECTRA-RCWA returns the value and the gradient in 0.23 s, and TorchRDIT at the series order of its paper, with errors of the same order, in 10.9 s (Figure~\ref{fig:compare}c). At \SI{1}{\micro\metre} the one-slice series is wrong by more than 100, and with slices of \SI{0.25}{\micro\metre} TorchRDIT takes 19.4 s against 0.30 s, 65 times longer. The value alone is 13 and 23 times faster, so most of the gap lies in the gradient. TorchRDIT differentiates through a graph that grows with the series order, the number of slices and the number of solves. Within 80\% of the card it could take only four of the 16 incidence directions per call at \SI{0.25}{\micro\metre}, and two at \SI{1}{\micro\metre}, and it still reached 18.9 and 14.0 GiB. SPECTRA-RCWA ran all 144 solves in one call with 6.0 and 8.4 GiB (Figure~\ref{fig:compare}d and Table S10). Thickness also enters the two codes differently: doubling adds work in proportion to the logarithm of the thickness, slices in proportion to the thickness itself.

\subsection{Speed on a Graphics Processor}

On a graphics processor, one evaluation of a 144-solve objective takes a quarter of a second. The benchmark is one evaluation of an imaging objective of a colour-router pixel under a camera lens:\cite{park2026imaging} 16 rays of the lens pupil and 9 wavelengths, 144 solves for both polarisations, with 17 $\times$ 17 Fourier orders ($578\times578$ layer matrices), a silicon nitride design layer \SI{0.8}{\micro\metre} thick with a period of \SI{2}{\micro\metre}, and a gradient with respect to $128\times128$ design values. As baselines we ran S4,\cite{liu2012s4} which gives the value on a processor, torcwa,\cite{kim2023torcwa} which is differentiable and runs on the graphics processor one solve at a time, and a code of ours that batches the eigendecomposition on the graphics processor. All graphics-processor codes ran in single precision on one RTX 5090 card.

\begin{figure}
\centering
\includegraphics[max width=\textwidth]{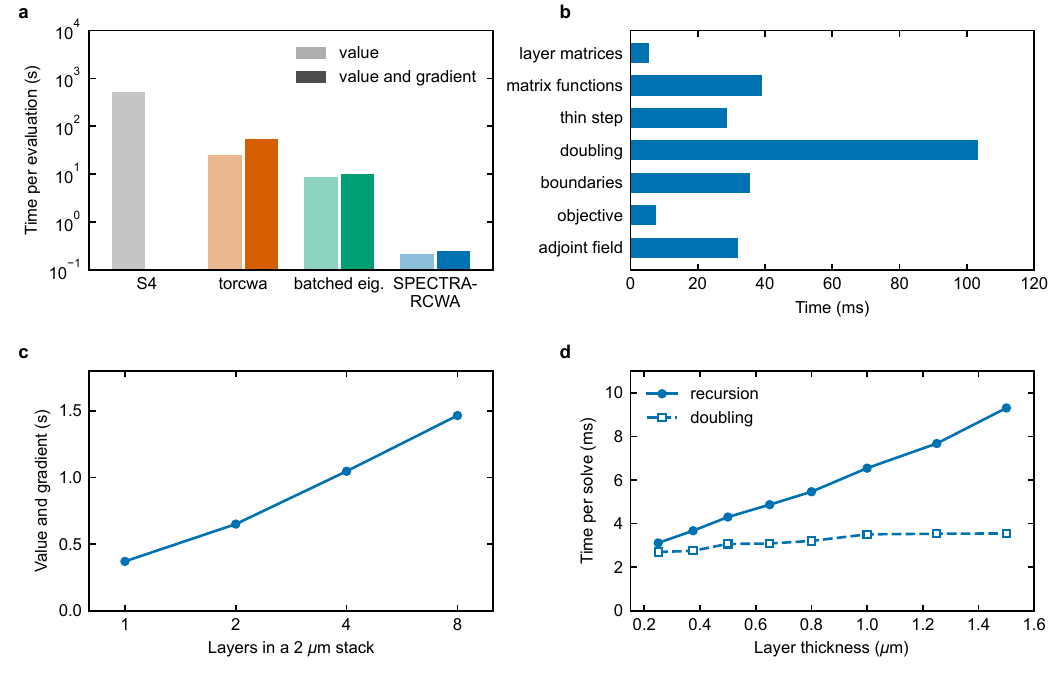}
\caption{\textbf{All 144 solves of an objective return a value and a gradient in a quarter of a second.} Speed of the 144-solve benchmark on one RTX 5090 card in single precision. (a) Time per evaluation for the value (light bars) and for the value and the gradient (dark bars) with S4 on 12 processor cores, torcwa, a batched eigendecomposition code and SPECTRA-RCWA. (b) Time of the stages of one SPECTRA-RCWA evaluation. (c) Time for the value and the gradient of a stack with a total thickness of \SI{2}{\micro\metre} against the number of design layers. (d) Time per solve of one layer against its thickness, with the admittance recursion and with doubling.}
\label{fig:speed}
\end{figure}

SPECTRA-RCWA returns the value in 0.217 s and the value with its gradient in 0.252 s, or 1.75 ms per solve (Figure~\ref{fig:speed}a and Table~\ref{tab:bench}). The same evaluation takes torcwa 54.8 s and the batched eigendecomposition code 10.2 s, 218 and 41 times longer. S4 computes the value alone in 529 s on 12 processor cores. Most of the gain comes from batching. A single solve of SPECTRA-RCWA with its gradient takes 34 ms for both polarisations, against 95 ms per polarisation for torcwa, and placing 144 solves in one batch lowers the cost per solve by a further factor of 19. The gradient adds 32 ms, or 13\% of the evaluation, to a forward pass of 220 ms, in which the doubling of the layer scattering matrix takes 103 ms and the matrix functions 39 ms (Figure~\ref{fig:speed}b).

\begin{table}
\caption{One evaluation of the benchmark (144 solves, both polarisations, $578\times578$ layer matrices)}
\label{tab:bench}
\begin{tabular}{lcccc}
\toprule
 & S4 & torcwa & batched eig. & SPECTRA-RCWA \\
\midrule
hardware & 12 cores & RTX 5090 & RTX 5090 & RTX 5090 \\
value (s) & 529 & 25.5 & 8.85 & 0.217 \\
value and gradient (s) & & 54.8 & 10.2 & 0.252 \\
peak memory (GiB) & & 4.8 & 4.2 & 6.6 \\
output error, grey & & $8.9\times10^{-6}$ & $6.0\times10^{-6}$ & $3.2\times10^{-6}$ \\
gradient error, grey & & $5.9\times10^{-3}$ & $1.8\times10^{-3}$ & $2.7\times10^{-6}$ \\
gradient error, binary & & $7.2\times10^{-2}$ & $5.4\times10^{-2}$ & $8.2\times10^{-6}$ \\
gradient error, at cutoff & & 0.49 & 2.0 & $3.8\times10^{-5}$ \\
\bottomrule
\end{tabular}
\end{table}

Stacks and thick layers scale as the formulation predicts. A stack with a total thickness of \SI{2}{\micro\metre} takes 0.37, 0.65, 1.05 and 1.47 s for the value and the gradient when it is divided into one, two, four and eight design layers, all 144 solves in one batch (Figure~\ref{fig:speed}c). For a single layer the time per solve of the recursion grows linearly with the thickness, from 3.1 ms at \SI{0.25}{\micro\metre} to 9.3 ms at \SI{1.5}{\micro\metre}, while doubling grows logarithmically, from 2.7 ms to 3.6 ms (Figure~\ref{fig:speed}d). A silicon nitride layer \SI{3}{\micro\metre} thick costs 4.0 ms per solve with doubling, and an absorbing silicon layer of equal thickness costs 4.1 ms, with gradient errors below $8\times10^{-5}$ (Table S5). On an RTX 4090 card the benchmark evaluation takes 0.391 s.

\section{Design Example: A Freeform Metagrating}

A freeform metagrating designed on a desktop processor deflects both polarisations by 65$^\circ$. The task is to send normally incident light of both polarisations into the first transmitted order at 65$^\circ$, a large angle for which freeform patterns are known to outperform libraries of simple shapes.\cite{sell2017large,phan2019high,jiang2019global} The layer has a refractive index of 2.4 and a thickness of \SI{0.6}{\micro\metre} on glass, with a cell of 698 nm by 317 nm. The objective is the first-order efficiency averaged over 21 wavelengths from 613 nm to 653 nm and the two polarisations, and each evaluation is one batch of 21 solves with $13\times5$ orders. The 2048 design values pass through a smoothing filter and a projection whose sharpness is raised in stages until the pattern is binary.\cite{jensen2011topology,christiansen2021inverse}

\begin{figure}
\centering
\includegraphics[max width=\textwidth]{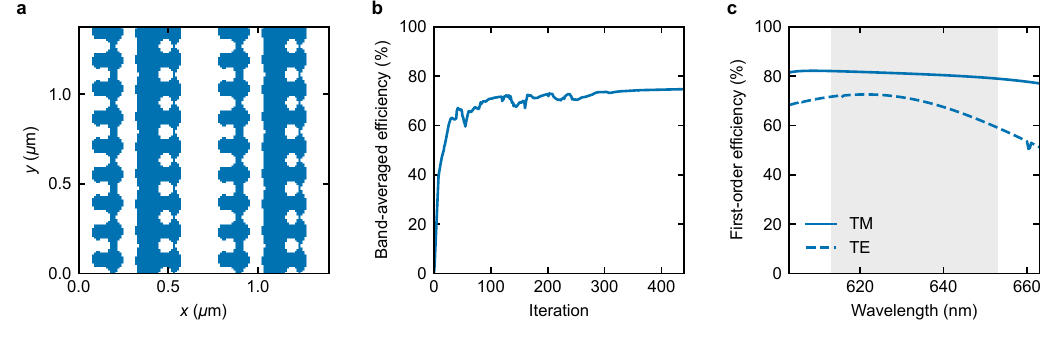}
\caption{\textbf{A freeform metagrating deflects both polarisations by 65$^\circ$.} (a) Binary pattern of the layer (index 2.4 in colour, air in white), with two periods along the direction of deflection, $x$. (b) Efficiency of the first transmitted order, averaged over the band and the two polarisations, during the optimisation. (c) Spectrum of the first-order efficiency of the final pattern for the two polarisations; the shaded band is the design band.}
\label{fig:design}
\end{figure}

The optimisation ran in double precision on a desktop processor and finished its 440 iterations in 3.4 min, 0.45 s per iteration. The binary pattern reaches a band-averaged efficiency of 74.7\%, with 81\% for the polarisation in the plane of diffraction and 69\% for the other (Figure~\ref{fig:design}). Increasing the truncation to $25\times11$ orders changes the band average to 74.4\%, and torcwa reproduces the efficiencies of the final pattern to $2\times10^{-12}$ at every truncation (Table S11). The adjoint gradient of this objective agrees with reverse-mode differentiation to $1.5\times10^{-15}$. A design problem with a spectral band and two polarisations is an interactive computation on a processor, and the same code moves it to a graphics processor without change when the cell or the number of orders grows.

\section{Conclusion}

We have shown that RCWA can keep the accuracy of the modal method without the eigendecomposition of its layer matrices. Three entire matrix functions replace the eigendecomposition, an admittance recursion or doubling replaces the modal scattering matrix, and one adjoint field replaces the computational graph. Each replacement keeps the accuracy of the modal method, to rounding error in double precision and to about $10^{-6}$ in single precision, and removes one obstacle to running an entire objective as a batch on a graphics processor: 144 solves with $578\times578$ matrices return a value and a gradient in a quarter of a second.

Nothing in the formulation is singular at a cutoff or at a degenerate pair of modes. The gradient keeps its accuracy at the points where eigendecomposition gradients are wrong by order one, and these points occur routinely in multilayer and multiwavelength designs. The work per step is chosen from the layer itself, and one solver covers a thin metasurface and a layer of 79 wavelengths with no setting to tune, where a series of fixed order must be sliced by hand and fails without a sign in the energy balance when it is not. The gradient comes from one adjoint field and not from a graph of the forward pass, so its memory does not grow with the series or the slices, and at matched accuracy it is 47 to 65 times faster than that of TorchRDIT on the same card.

Objectives that were too costly for gradient-based design with a full-wave solver come within reach: image-sensor pixels optimised under the full pupil of a camera lens, stacks of many patterned layers, and devices specified over continuous bands of wavelength and angle. The stored scattering matrices of fixed layers could substantially shorten the design of patterned layers above detectors and multilayer coatings.

The approach is not specific to RCWA. It needs only a first-order system that is propagated through a uniform layer, and any Fourier-space method of this kind could replace its modes by entire functions in the same way.

\section{Methods}

SPECTRA-RCWA is written in PyTorch and runs on processors and on graphics processors in single or double precision. Permittivities enter by the Laurent rule with the inverse of the permittivity matrix in $\mat P$, as in torcwa, and all Fourier orders of a rectangular truncation are kept. Every solve runs at a frequency raised by a relative $10^{-4}$, which keeps the orders of the half-spaces off their Rayleigh points. Solver settings are listed in Table S1.

Graphics-processor measurements were taken on an RTX 5090 card (32 GB) with PyTorch 2.10.0; each time is the fastest of at least three evaluations after a warm-up call. The accuracy reference of the benchmark is a double-precision solution with a converged matrix square root. The grey, binary and cutoff layers of Figure~\ref{fig:gradient}c and Table~\ref{tab:bench} are a design layer with intermediate densities, the same layer after binarisation, and a \SI{0.25}{\micro\metre} layer of a multilayer design that contains a solve with an order at cutoff. torcwa 0.1.4.2 solved the 288 single-polarisation problems one by one, with a checkpoint per wavelength for the gradient. S4 was run in 12 processes on an Intel i7-12700 processor. The batched eigendecomposition code diagonalises 48 layer matrices at a time. Details are given in Section S3 of the Supplementary Information.

The computations of Figures~\ref{fig:accuracy}, \ref{fig:gradient}a,b, \ref{fig:compare}a,b and \ref{fig:design} ran on an Intel i7-12700 desktop processor. TorchRDIT 0.2.0 was installed from its public release; its series order of 15 is the one used in its paper and 20 the largest reported there. Timings on the processor are medians of nine runs; the comparison of Figure~\ref{fig:compare}c,d on the graphics card gives medians of five runs (Section S6). The metagrating was optimised with the Adam optimiser, a Gaussian density filter of 2.5 grid cells and a hyperbolic-tangent projection with a sharpness raised from 2 to 64 in 11 levels of 40 iterations.

\section*{Supplementary Information}
The Supplementary Information after the references gives the layer matrices and their structure, the derivation of the adjoint field, the solver settings, the benchmark protocol and additional measurements.

\section*{Data and Code Availability}

The code is available at \url{https://github.com/hyoseokp/spectra-rcwa}. The measurements behind every figure and table are stored with the benchmark scripts of the repository and with the figure scripts of this article.

\section*{Competing Interests}

The author declares no competing financial interest.

\bibliographystyle{unsrtnat}
\bibliography{refs}

\clearpage
\setcounter{section}{0}\setcounter{figure}{0}\setcounter{table}{0}\setcounter{equation}{0}
\renewcommand{\thesection}{S\arabic{section}}\renewcommand{\thefigure}{S\arabic{figure}}
\renewcommand{\thetable}{S\arabic{table}}\renewcommand{\theequation}{S\arabic{equation}}
\begin{center}{\Large\bfseries Supplementary Information}\end{center}

\section{Layer Matrices}

The fields are periodic in $x$ and $y$ and are expanded in $N$ Fourier orders with transverse wavevectors $(k_{x,n},k_{y,n})$ in units of $\omega=2\pi/\lambda$; $\mat K_x$ and $\mat K_y$ are the diagonal matrices of these components. The time dependence is $\exp(-i\omega t)$ and $z$ points from the incidence medium to the substrate. The vectors $\mat e$ and $\mat h$ of eq~1 of the main text collect the Fourier amplitudes of the two tangential components of the electric and of the magnetic field, with the magnetic field in units that absorb the vacuum impedance. The permittivity of a layer enters through its Fourier (Toeplitz) matrix $\mat E$, with entries $\mat E_{mn}=\hat\varepsilon_{m-n}$, by the Laurent rule,\cite{li1996fourier,li1997new} as in torcwa.\cite{kim2023torcwa} In block form
\begin{equation}
\mat P=\begin{pmatrix}\mat K_x\mat E^{-1}\mat K_y & \mat I-\mat K_x\mat E^{-1}\mat K_x\\ \mat K_y\mat E^{-1}\mat K_y-\mat I & -\mat K_y\mat E^{-1}\mat K_x\end{pmatrix},\qquad
\mat Q=\begin{pmatrix}-\mat K_x\mat K_y & \mat K_x^2-\mat E\\ \mat E-\mat K_y^2 & \mat K_y\mat K_x\end{pmatrix}.
\end{equation}
Both are a fixed sparse part plus a term of rank $N$. With
\begin{equation}
\mat K_L=\begin{pmatrix}\mat K_x\\ \mat K_y\end{pmatrix},\qquad \mat K_R=\begin{pmatrix}\mat K_y & -\mat K_x\end{pmatrix},\qquad
\mat J=\begin{pmatrix}0&\mat I\\-\mat I&0\end{pmatrix},\qquad \hat{\mat E}=\begin{pmatrix}0&-\mat E\\ \mat E&0\end{pmatrix},
\end{equation}
they read $\mat P=\mat J+\mat K_L\mat E^{-1}\mat K_R$ and $\mat Q=-\mat K_L\mat K_R+\hat{\mat E}$. Since $\mat K_R\mat K_L=0$,
\begin{equation}
\mat A=\mat P\mat Q=\begin{pmatrix}\mat E&0\\0&\mat E\end{pmatrix}-\mat J\mat K_L\mat K_R-\mat K_L\,\mat E^{-1}\begin{pmatrix}\mat K_x\mat E&\mat K_y\mat E\end{pmatrix},
\end{equation}
which needs the two $N\times N$ products $\mat E^{-1}(\mat K_x\mat E)$ and $\mat E^{-1}(\mat K_y\mat E)$ and elementwise operations. A product of $\mat P$ with a matrix costs one $N$-row product, a quarter of a dense $2N\times2N$ product, and a product with $\mat Q$ costs half of one. The homogeneous media above and below the stack have diagonal blocks, and their admittances $\mat V_i$ and $\mat V_o$, defined by $\mat h=\mat V\mat e$ for a wave travelling toward $+z$, are known in closed form.

A pattern is described by a density $\rho(x,y)\in[0,1]$ on a grid, and the permittivity is $\varepsilon=\varepsilon_{\rm fill}+\rho\,(\varepsilon_{\rm design}-\varepsilon_{\rm fill})$. A fast Fourier transform of $\varepsilon$ gives $\mat E$ for all wavelengths of the batch.

\section{Adjoint Field}

Let the design change $\mat P$ and $\mat Q$ of a layer by $\delta\mat P$ and $\delta\mat Q$. The variation of eq~1 of the main text is
\begin{equation}
\frac{d\,\delta\mat e}{dz}=i\omega(\mat P\,\delta\mat h+\delta\mat P\,\mat h),\qquad
\frac{d\,\delta\mat h}{dz}=i\omega(\mat Q\,\delta\mat e+\delta\mat Q\,\mat e).
\end{equation}
For any pair $(\mat a,\mat b)$ that obeys $d\mat a/dz=i\omega\mat Q^{H}\mat b$ and $d\mat b/dz=i\omega\mat P^{H}\mat a$, the terms with $\delta\mat e$ and $\delta\mat h$ cancel in
\begin{equation}
\frac{d}{dz}\left(\mat a^{H}\delta\mat e+\mat b^{H}\delta\mat h\right)=i\omega\left(\mat a^{H}\delta\mat P\,\mat h+\mat b^{H}\delta\mat Q\,\mat e\right).
\label{eq:bilinear}
\end{equation}
The bilinear form on the left is continuous across the interfaces, because the tangential fields are. Integrating eq~\ref{eq:bilinear} from the top of the stack, $z=0$, to its bottom, $z=d$, leaves two boundary terms. At the top the field is the fixed incident field plus a reflected field $\mat r$, so $\delta\mat e=\delta\mat r$ and $\delta\mat h=-\mat V_i\,\delta\mat r$, and the term $(\mat a-\mat V_i^{H}\mat b)^{H}\delta\mat r$ vanishes when $\mat a=\mat V_i^{H}\mat b$ at $z=0$. At the bottom the field is outgoing, $\delta\mat h=\mat V_o\,\delta\mat e$, and the term is $(\mat a+\mat V_o^{H}\mat b)^{H}\delta\mat e(d)$. For a real objective $J$ of the transmitted field with $\delta J=\operatorname{Re}[\mat g^{H}\delta\mat e(d)]$, the condition $\mat a+\mat V_o^{H}\mat b=\mat g$ at $z=d$ gives eq~6 of the main text,
\begin{equation}
\delta J=\operatorname{Re}\int_0^d i\omega\left(\mat a^{H}\delta\mat P\,\mat h+\mat b^{H}\delta\mat Q\,\mat e\right)dz .
\end{equation}
An objective that also depends on the reflected field adds a source term to the condition at the top in the same way. Both $\delta\mat P$ and $\delta\mat Q$ are linear in $\delta\mat E$ and $\delta(\mat E^{-1})$ (Section S1), so the integral is accumulated as two $N\times N$ matrices, the derivatives of $J$ with respect to $\mat E$ and to $\mat E^{-1}$, and these are mapped to the density through the Fourier transform of the pattern.

The adjoint system has the structure of the forward one with $\mat P$ and $\mat Q$ replaced by $\mat Q^{H}$ and $\mat P^{H}$, so its admittance recursion uses the conjugate transposes of the factors that the forward pass has saved. The forward field is carried downward and the adjoint field upward through these factors, each in its stable direction, for two columns per solve. Within a step of thickness $h$ the fields at a node $z_0+t$ follow from those on the nearer plane by the Taylor series of the step propagator in $t$, which converges in a few terms; the series is cut when its next term falls below the tolerance of Table~\ref{tab:settings}.

\section{Solver Settings and Benchmark Protocol}

Table~\ref{tab:settings} lists the settings of the solver, which are the same for every result of the article. In single precision the polynomial degree and the scaling bound are chosen so that the truncation error of the polynomial stays below the rounding error of the products.

\begin{table}
\caption{Solver settings}
\label{tab:settings}
\begin{tabular}{lcc}
\toprule
 & single precision & double precision \\
\midrule
step bound, $\omega h k_{\max}$ & 4.5 & 4.5 \\
polynomial degree of $\mat C-\mat I$ and $\mat S$ & 6 & 8 \\
bound on the norm of the scaled argument & 6 & 1 \\
doubling instead of the recursion, from & 7 steps & 7 steps \\
permittivity of the reference medium & $4+4i$ & $4+4i$ \\
Gauss--Legendre nodes per step, gradient & 6 & 16 \\
first omitted Taylor term, gradient & $10^{-5}$ & $10^{-16}$ \\
relative frequency shift & $10^{-4}$ & $10^{-4}$ \\
\bottomrule
\end{tabular}
\end{table}

The scaling exponent $s$ is the smallest positive integer with $\max\big(\|\mat B^2\|_1^{1/2},\|\mat B^3\|_1^{1/3}\big)\le\theta\,4^{s}$, where $\mat B=(\omega h)^2\mat A$ and $\theta$ is the bound of Table~\ref{tab:settings}; one value serves the whole batch. With $\mat X=\mat C-\mat I$ of the scaled argument, a double-angle step is $\mat S\leftarrow\mat S+\mat S\mat X$ followed by $\mat X\leftarrow4\mat X+2\mat X^2$, and $\mat G$ is formed as $-\tfrac12\mat S^2$ before the last step. In single precision the evaluation costs $5+2s$ matrix products.

Table~\ref{tab:protocol} gives the protocol of the benchmark on the graphics processor. The accuracy reference solves one ray at a time in double precision with a matrix square root from 14 Denman--Beavers iterations and the modal scattering-matrix formulation; in double precision SPECTRA-RCWA reproduces the objective of this reference to $7\times10^{-15}$. The output error is the largest relative error of the detector powers, and the gradient error is the relative $L_2$ difference of the gradients with respect to the $128\times128$ design values.

\begin{table}
\caption{Protocol of the benchmark}
\label{tab:protocol}
\begin{tabular}{lp{11cm}}
\toprule
problem & 16 pupil rays (numerical aperture 0.30, chief ray at 2.5$^\circ$) $\times$ 9 wavelengths, both polarisations; $17\times17$ orders; silicon nitride layer of \SI{0.8}{\micro\metre} on oxide, period \SI{2}{\micro\metre}; $128\times128$ design values \\
graphics processor & RTX 5090 (32 GB), PyTorch 2.10.0, single precision; second card RTX 4090 (24 GB) \\
timing & fastest of at least three evaluations after one warm-up call \\
SPECTRA-RCWA & release 1.1.0; all 144 solves in one batch \\
torcwa & version 0.1.4.2; 288 single-polarisation solves one by one; gradient with one checkpoint per wavelength (the graph of all solves would need 41 GiB) \\
batched eigendecomposition & 48 layer matrices per call on the graphics processor; reverse-mode gradient \\
S4 & 12 processes on an Intel i7-12700 processor; value only; transmission within $1.7\times10^{-4}$ of torcwa \\
\bottomrule
\end{tabular}
\end{table}

\section{Additional Measurements on the Graphics Processor}

Table~\ref{tab:errors} completes Figure~3c of the main text with the output errors and with the square-root formulation, which runs in the same code and batch as SPECTRA-RCWA with four Denman--Beavers iterations. Table~\ref{tab:stages} gives the stage times of Figure~5b, Table~\ref{tab:thick} the thick layers and Table~\ref{tab:stacks} the layer stacks on two cards. For the absorbing silicon layer of Table~\ref{tab:thick} the absorbed fractions of the two polarisations are 0.7149 and 0.7143 in single precision and in the reference alike. In double precision the thick-layer values agree with the reference to $6\times10^{-16}$ and the gradients to $7\times10^{-13}$.

\begin{table}
\caption{Errors of the benchmark evaluation against the double-precision reference for three design layers}
\label{tab:errors}
\begin{tabular}{llcccc}
\toprule
layer & quantity & SPECTRA-RCWA & square root & torcwa & batched eig. \\
\midrule
grey & output & $3.2\times10^{-6}$ & $6.8\times10^{-6}$ & $8.9\times10^{-6}$ & $6.0\times10^{-6}$ \\
grey & gradient & $2.7\times10^{-6}$ & $7.2\times10^{-4}$ & $5.9\times10^{-3}$ & $1.8\times10^{-3}$ \\
binary & output & $3.4\times10^{-6}$ & $6.3\times10^{-6}$ & $8.9\times10^{-6}$ & $9.5\times10^{-6}$ \\
binary & gradient & $8.2\times10^{-6}$ & $4.3\times10^{-3}$ & $7.2\times10^{-2}$ & $5.4\times10^{-2}$ \\
at cutoff & output & $1.2\times10^{-6}$ & $7.7\times10^{-6}$ & $6.2\times10^{-5}$ & $3.5\times10^{-5}$ \\
at cutoff & gradient & $3.8\times10^{-5}$ & $5.1\times10^{-1}$ & $4.9\times10^{-1}$ & $2.0\times10^{0}$ \\
 
\bottomrule
\end{tabular}
\end{table}

\begin{table}
\caption{Stage times of one benchmark evaluation of SPECTRA-RCWA (RTX 5090, medians)}
\label{tab:stages}
\begin{tabular}{lc}
\toprule
stage & time (ms) \\
\midrule
permittivity matrix and its inverse & 2.4 \\
layer matrix $\mathbf{A}$ & 3.1 \\
matrix functions $\mathbf{C}$, $\mathbf{S}$, $\mathbf{G}$ & 39.0 \\
thin step in the reference basis & 28.8 \\
doubling & 103.3 \\
boundaries and the two right-hand sides & 35.6 \\
detector field and objective & 7.5 \\
adjoint field on the planes & 14.0 \\
integral over the layer & 18.1 \\
total & 251.9 \\
 
\bottomrule
\end{tabular}
\end{table}

\begin{table}
\caption{Thick layers at $17\times17$ orders and 540 nm (RTX 5090, single precision; errors for one ray, time and memory per solve for 16 rays in one batch). The silicon layer has the permittivity $17.1+0.28i$ and a \SI{0.2}{\micro\metre} oxide trench}
\label{tab:thick}
\begin{tabular}{llccccc}
\toprule
layer, thickness (\si{\micro\metre}) & route & steps & value error & gradient error & ms & MiB \\
\midrule
SiN, 0.8 & recursion & 7 & $1.4\times10^{-8}$ & $3.2\times10^{-5}$ & 5.5 & 59 \\
SiN, 0.8 & doubling & 8 & $7.5\times10^{-8}$ & $9.1\times10^{-6}$ & 3.2 & 48 \\
SiN, 1.5 & recursion & 12 & $2.0\times10^{-8}$ & $3.3\times10^{-5}$ & 9.3 & 96 \\
SiN, 1.5 & doubling & 16 & $1.3\times10^{-9}$ & $1.2\times10^{-5}$ & 3.6 & 61 \\
SiN, 3 & recursion & 24 & $1.0\times10^{-7}$ & $5.8\times10^{-5}$ & 16.6 & 172 \\
SiN, 3 & doubling & 32 & $2.9\times10^{-8}$ & $2.9\times10^{-5}$ & 4.0 & 80 \\
Si with a trench, 3 & recursion & 24 & $1.7\times10^{-7}$ & $5.8\times10^{-5}$ & 16.8 & 172 \\
Si with a trench, 3 & doubling & 32 & $8.8\times10^{-9}$ & $7.7\times10^{-5}$ & 4.1 & 80 \\
 
\bottomrule
\end{tabular}
\end{table}

\begin{table}
\caption{Value and gradient for stacks with a total thickness of \SI{2}{\micro\metre}, 144 solves}
\label{tab:stacks}
\begin{tabular}{ccccccc}
\toprule
 & & \multicolumn{2}{c}{RTX 5090} & \multicolumn{3}{c}{RTX 4090} \\
layers & thickness (\si{\micro\metre}) & time (s) & memory (GiB) & time (s) & memory (GiB) & solves per batch \\
\midrule
1 & 2 & 0.37 & 10.8 & 0.50 & 10.8 & 144 \\
2 & 1 & 0.65 & 15.3 & 0.82 & 15.3 & 144 \\
4 & 0.5 & 1.05 & 20.2 & 3.17 & 13.0 & 48 \\
8 & 0.25 & 1.47 & 26.8 & 4.79 & 13.1 & 36 \\
 
\bottomrule
\end{tabular}
\end{table}

\section{Validation on a Processor}

\begin{figure}
\centering
\includegraphics[max width=\textwidth]{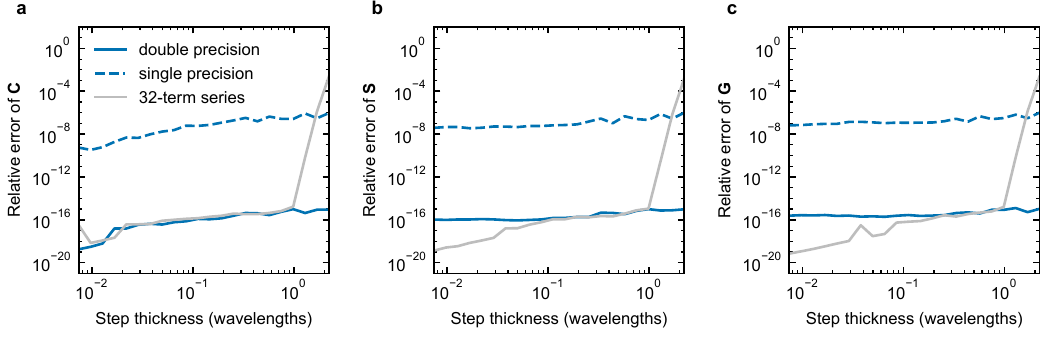}
\caption{Relative error of (a) $\mat C$, (b) $\mat S$ and (c) $\mat G$ against a 256-bit reference as a function of the step thickness, for the scaled evaluation in double and single precision and for a 32-term power series without scaling. Structure as in Figure~2a of the main text.}
\label{fig:matfun}
\end{figure}

\begin{figure}
\centering
\includegraphics[max width=\textwidth]{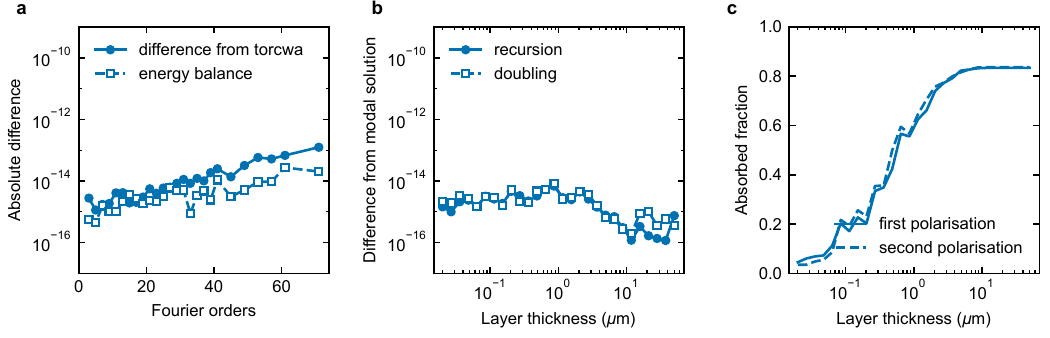}
\caption{(a) Largest difference between the diffraction efficiencies of SPECTRA-RCWA and of torcwa for the lamellar grating of Figure~2b of the main text, and deviation of the summed efficiencies from one. (b) Difference from the modal solution against the thickness for the freeform layer of Figure~2c with the absorbing permittivity $17.1+0.28i$. (c) Absorbed fraction of that layer for the two incident polarisations.}
\label{fig:grating}
\end{figure}

\begin{figure}
\centering
\includegraphics[max width=\textwidth]{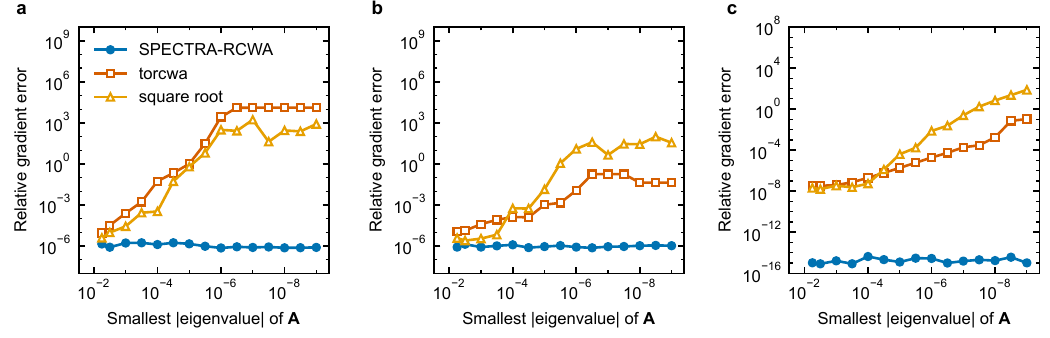}
\caption{Gradient error next to a cutoff, for the layer of Figure~3a,b of the main text. (a,b) Single precision, objective with random weights on all orders, at the cutoff where $\mat P$ becomes singular (a) and where $\mat Q$ does (b). (c) Double precision, zeroth-order transmission, at the first cutoff; the reference is the adjoint field in double precision and the curve of SPECTRA-RCWA is reverse-mode differentiation of the same forward pass.}
\label{fig:cutoff}
\end{figure}

Figure~\ref{fig:matfun} extends Figure~2a of the main text to the three matrix functions. Figure~\ref{fig:grating} gives the differences behind Figure~2b and the thickness scan of an absorbing layer. Figure~\ref{fig:cutoff} repeats the gradient scan of Figure~3a,b for a second objective, a sum of all diffraction efficiencies with random weights between $-1$ and 1, and shows the scan in double precision.

\section{Comparison with TorchRDIT}

The structure is a $16\times16$ pattern with densities 0.1 and 0.9 of a material of index 2.0 in air, on a substrate of permittivity 2.125, with a period of \SI{1.2}{\micro\metre} and $7\times7$ orders ($98\times98$ layer matrices). TorchRDIT 0.2.0 ran from its public release with PyTorch 2.6.0 on an Intel i7-12700 processor. The frequency shift of SPECTRA-RCWA was set to zero so that both codes solve at the same wavelengths. Each code is compared with a converged modal solution in its own conventions; the modal results of the two codes agree order by order to between $2\times10^{-10}$ and $2\times10^{-9}$. Table~\ref{tab:rditthick} gives the largest error of the diffraction efficiencies over four wavelengths (500, 570, 640 and 700 nm), two directions of incidence and both polarisations. Table~\ref{tab:rditgrad} gives the gradient of the zeroth-order transmission with respect to the 256 densities for a layer of \SI{0.3}{\micro\metre}, and Table~\ref{tab:rdittime} the times for a batch of nine wavelengths from 500 to 700 nm.

\begin{table}
\caption{Largest error of the diffraction efficiencies against the layer thickness $d$. Columns 3 to 6: TorchRDIT in single precision with one slice and series order 10, 15 and 20, and with order 20 and slices of at most \SI{0.25}{\micro\metre}; last column: the latter in double precision}
\label{tab:rditthick}
\footnotesize
\setlength{\tabcolsep}{3pt}
\begin{tabular}{cccccccc}
\toprule
 & \multicolumn{2}{c}{SPECTRA-RCWA} & \multicolumn{5}{c}{TorchRDIT} \\
$d$ (\si{\micro\metre}) & single & double & order 10 & order 15 & order 20 & sliced & sliced, double \\
\midrule
0.05 & $1.1\times10^{-6}$ & $2.4\times10^{-15}$ & $8.4\times10^{-7}$ & $8.4\times10^{-7}$ & $8.4\times10^{-7}$ & $8.4\times10^{-7}$ & $2.1\times10^{-10}$ \\
0.075 & $1.1\times10^{-6}$ & $5.1\times10^{-15}$ & $8.2\times10^{-7}$ & $8.2\times10^{-7}$ & $8.2\times10^{-7}$ & $8.2\times10^{-7}$ & $2.1\times10^{-10}$ \\
0.1 & $7.9\times10^{-7}$ & $3.4\times10^{-15}$ & $8.4\times10^{-7}$ & $7.3\times10^{-7}$ & $7.3\times10^{-7}$ & $7.3\times10^{-7}$ & $2.2\times10^{-10}$ \\
0.15 & $1.4\times10^{-6}$ & $7.5\times10^{-15}$ & $7.2\times10^{-7}$ & $8.7\times10^{-7}$ & $8.7\times10^{-7}$ & $8.7\times10^{-7}$ & $3.7\times10^{-10}$ \\
0.2 & $5.7\times10^{-7}$ & $5.6\times10^{-15}$ & $1.1\times10^{-4}$ & $6.1\times10^{-7}$ & $6.2\times10^{-7}$ & $6.2\times10^{-7}$ & $3.8\times10^{-10}$ \\
0.3 & $7.8\times10^{-7}$ & $4.1\times10^{-15}$ & $9.0\times10^{-3}$ & $1.3\times10^{-6}$ & $5.1\times10^{-7}$ & $3.2\times10^{-5}$ & $3.3\times10^{-10}$ \\
0.4 & $3.5\times10^{-7}$ & $3.1\times10^{-15}$ & $4.1\times10^{-2}$ & $5.0\times10^{-4}$ & $8.6\times10^{-7}$ & $1.0\times10^{-5}$ & $6.4\times10^{-10}$ \\
0.5 & $5.4\times10^{-7}$ & $4.9\times10^{-15}$ & $1.9\times10^{-1}$ & $2.9\times10^{-2}$ & $9.3\times10^{-5}$ & $4.8\times10^{-6}$ & $5.0\times10^{-10}$ \\
0.75 & $7.4\times10^{-7}$ & $6.4\times10^{-15}$ & $3.1\times10^{-1}$ & $3.2\times10^{-1}$ & $1.7\times10^{-1}$ & $2.7\times10^{-6}$ & $5.2\times10^{-10}$ \\
1 & $1.0\times10^{-6}$ & $4.2\times10^{-15}$ & $8.2\times10^{-1}$ & $5.9\times10^{-1}$ & $1.9\times10^{-1}$ & $5.1\times10^{-6}$ & $3.8\times10^{-10}$ \\
1.5 & $2.7\times10^{-6}$ & $1.4\times10^{-14}$ & $8.3\times10^{-1}$ & $8.1\times10^{-1}$ & $7.0\times10^{-1}$ & $4.0\times10^{-6}$ & $1.2\times10^{-9}$ \\
2 & $1.8\times10^{-6}$ & $1.5\times10^{-14}$ & $3.3\times10^{0}$ & $7.9\times10^{2}$ & $1.6\times10^{3}$ & $3.6\times10^{-6}$ & $1.1\times10^{-9}$ \\
3 & $3.8\times10^{-6}$ & $2.0\times10^{-14}$ & $1.1\times10^{2}$ & $1.3\times10^{3}$ & $4.3\times10^{3}$ & $7.5\times10^{-6}$ & $1.2\times10^{-9}$ \\
 
\bottomrule
\end{tabular}
\end{table}

\begin{table}
\caption{Relative error of the single-precision gradient next to a cutoff}
\label{tab:rditgrad}
\begin{tabular}{lcccc}
\toprule
 & smallest & & TorchRDIT, & TorchRDIT, \\
case & $|$eigenvalue$|$ & SPECTRA-RCWA & series, order 20 & eigendecomposition \\
\midrule
away from cutoff & $1.9\times10^{-2}$ & $9.9\times10^{-7}$ & $1.7\times10^{-6}$ & $5.8\times10^{-3}$ \\
$\mathbf{P}$ singular & $1.0\times10^{-4}$ & $7.8\times10^{-7}$ & $1.1\times10^{-6}$ & $2.6\times10^{0}$ \\
$\mathbf{P}$ singular & $1.0\times10^{-6}$ & $9.6\times10^{-7}$ & $1.2\times10^{-6}$ & $1.8\times10^{2}$ \\
$\mathbf{P}$ singular & $1.0\times10^{-8}$ & $8.7\times10^{-7}$ & $7.6\times10^{-7}$ & $7.2\times10^{2}$ \\
$\mathbf{P}$ singular & $1.0\times10^{-10}$ & $9.9\times10^{-7}$ & $7.7\times10^{-7}$ & $7.2\times10^{2}$ \\
$\mathbf{Q}$ singular & $1.0\times10^{-4}$ & $6.6\times10^{-7}$ & $1.6\times10^{-6}$ & $5.6\times10^{-1}$ \\
$\mathbf{Q}$ singular & $1.0\times10^{-6}$ & $7.4\times10^{-7}$ & $9.8\times10^{-7}$ & $2.1\times10^{2}$ \\
$\mathbf{Q}$ singular & $1.0\times10^{-8}$ & $8.3\times10^{-7}$ & $1.1\times10^{-6}$ & $2.6\times10^{3}$ \\
$\mathbf{Q}$ singular & $1.0\times10^{-10}$ & $8.4\times10^{-7}$ & $1.1\times10^{-6}$ & $2.6\times10^{3}$ \\
 
\bottomrule
\end{tabular}
\end{table}

\begin{table}
\caption{Time on the processor for a batch of nine wavelengths and both polarisations in single precision (medians of nine runs)}
\label{tab:rdittime}
\begin{tabular}{clcc}
\toprule
thickness (\si{\micro\metre}) & code and setting & value (ms) & value and gradient (ms) \\
\midrule
0.25 & SPECTRA-RCWA & 16 & 38 \\
0.25 & TorchRDIT, order 15, one slice & 95 & 260 \\
0.25 & TorchRDIT, eigendecomposition mode & 167 & 362 \\
1 & SPECTRA-RCWA & 29 & 62 \\
1 & TorchRDIT, order 20, four slices & 123 & 729 \\
1 & TorchRDIT, eigendecomposition mode & 169 & 348 \\
 
\bottomrule
\end{tabular}
\end{table}

On the graphics card, both codes solved a problem of the size of the benchmark: $8\times8$ orders ($578\times578$ layer matrices), a grey pattern with densities between 0.1 and 0.9 of a material of permittivity 4.0 on a $64\times64$ grid, a period of \SI{1.2}{\micro\metre}, a substrate of permittivity 2.125, nine wavelengths from 500 to 700 nm and 16 incidence directions (polar angles of 3, 7, 11 and 15$^\circ$ at azimuths of 20, 110, 200 and 290$^\circ$), 144 solves with both polarisations. Each code ran on one RTX 5090 card with PyTorch 2.10.0 in single precision; the times are medians of five runs after two warm-up calls. Value errors are relative to an eigendecomposition solution in double precision, gradient errors to the double-precision gradient. Each code could use at most 80\% of the card; a setting that did not fit was run in calls over fewer incidence directions, and its time is the sum over the calls. The eigendecomposition mode of TorchRDIT was not timed on the card, because PyTorch computes the nonsymmetric eigendecomposition of a tensor on the graphics card on the host processor. Table~\ref{tab:rditgpu} lists the results.

\begin{table}
\caption{Comparison on one RTX 5090 card, 144 solves in single precision: time for the value and for the value and the gradient, peak memory of the latter, incidence directions per call (of 16), and errors of the value and of the gradient}
\label{tab:rditgpu}
\scriptsize
\setlength{\tabcolsep}{2.5pt}
\begin{tabular}{clcccccc}
\toprule
$d$ (\si{\micro\metre}) & code and setting & value (s) & + gradient (s) & memory (GiB) & directions & value error & gradient error \\
\midrule
0.25 & SPECTRA-RCWA & 0.21 & 0.23 & 6.0 & 16 & $3.6\times10^{-6}$ & $5.0\times10^{-7}$ \\
0.25 & TorchRDIT, order 15, one slice & 2.68 & 10.94 & 18.9 & 4 & $8.4\times10^{-6}$ & $1.4\times10^{-5}$ \\
0.25 & TorchRDIT, order 20, one slice & 2.85 & 11.53 & 20.7 & 4 & $9.0\times10^{-6}$ & $1.6\times10^{-5}$ \\
1 & SPECTRA-RCWA & 0.25 & 0.30 & 8.4 & 16 & $5.2\times10^{-6}$ & $2.4\times10^{-6}$ \\
1 & TorchRDIT, order 15, one slice & 2.71 & 11.57 & 18.9 & 4 & $1.1\times10^{2}$ & $7.9\times10^{4}$ \\
1 & TorchRDIT, order 20, one slice & 2.86 & 11.29 & 20.7 & 4 & $4.1\times10^{2}$ & $2.6\times10^{5}$ \\
1 & TorchRDIT, order 20, four slices & 5.74 & 19.43 & 14.0 & 2 & $3.7\times10^{-5}$ & $4.0\times10^{-5}$ \\
 
\bottomrule
\end{tabular}
\end{table}

\section{Metagrating}

\begin{figure}
\centering
\includegraphics[max width=\textwidth]{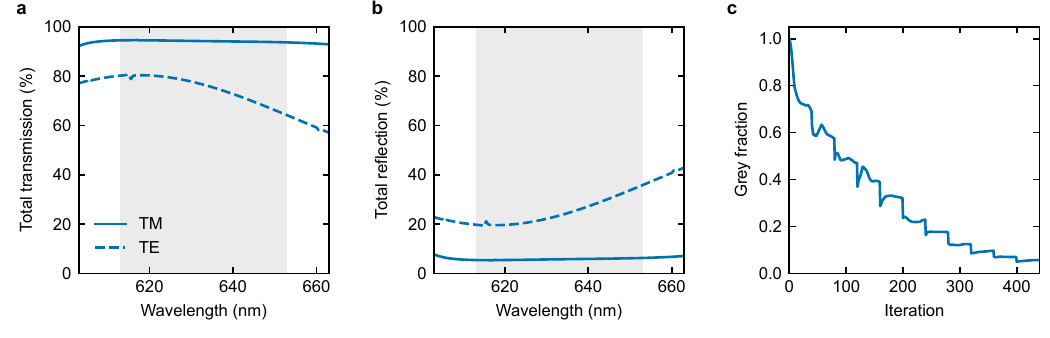}
\caption{Metagrating of Figure~6 of the main text. (a) Total transmission and (b) total reflection of the final pattern for the two polarisations; the shaded band is the design band. (c) Mean of $4\rho(1-\rho)$ over the design grid, which is one for a uniform grey pattern and zero for a binary one, during the optimisation.}
\label{figS:design}
\end{figure}

The cell measures 698.4 nm along $x$, which places the first order at 65$^\circ$ in air at 633 nm, and 316.5 nm along $y$, for which only the zeroth order propagates in that direction. Light is incident from the glass substrate (index 1.46) at normal incidence. The design grid has $64\times32$ values. The objective is the mean of the first-order transmission over the 21 wavelengths and the two polarisations. The optimiser step is $\min(0.03,\,0.12/\beta)$ for the projection sharpness $\beta$. The final binary pattern transmits 81.0\% and 70.6\% of the two polarisations into the first order at 633 nm. Table~\ref{tab:design} gives the band-averaged efficiency of this pattern at larger truncations, and Figure~\ref{figS:design} its total transmission and reflection and the binarisation during the optimisation. Release 1.1.0 of the code reproduces the stored efficiencies of the pattern to all digits in double precision.

\begin{table}
\caption{Final metagrating pattern at four truncations, double precision}
\label{tab:design}
\begin{tabular}{ccccc}
\toprule
orders & matrix size & band-averaged efficiency (\%) & difference from torcwa & energy balance \\
\midrule
$13\times5$ & 130 & 74.73 & $3.2\times10^{-14}$ & $2.2\times10^{-14}$ \\
$17\times7$ & 238 & 74.45 & $4.0\times10^{-14}$ & $1.4\times10^{-14}$ \\
$21\times9$ & 378 & 74.44 & $6.9\times10^{-13}$ & $2.7\times10^{-14}$ \\
$25\times11$ & 550 & 74.44 & $1.5\times10^{-12}$ & $3.9\times10^{-14}$ \\
 
\bottomrule
\end{tabular}
\end{table}

\end{document}